\pdfoutput=1

\documentclass[reprint,
superscriptaddress,
 amsmath,amssymb,
 aps,
longbibliography,
prl
]{revtex4-2}

\usepackage{xparse}

\usepackage{appendix}
\usepackage{graphicx}% Include figure files
\graphicspath{ {./Figures/} }

\usepackage{dcolumn}% Align table columns on decimal point
\usepackage{bm}% bold math
\usepackage{hyperref}% add hypertext capabilities
\usepackage{xcolor} % added by the author

\usepackage{comment}
\usepackage{dsfont}

\newcommand{\ket}[1]{\ensuremath{\left| #1 \right>}}
\newcommand{\bra}[1]{\ensuremath{\left< #1 \right|}}
\newcommand{\braket}[2]{\ensuremath{\left< #1 \ \vphantom{#2} \right| 
\left. #2 \vphantom{#1} \right>}}

\newcommand{\Tr}{\text{Tr}}

\usepackage{slashed}

\newcommand{\Lim}[1]{\raisebox{0.5ex}{\scalebox{0.8}{$\displaystyle \lim_{#1}\;$}}}
\usepackage{bm}
\newcommand{\vect}[1]{{\boldsymbol{\mathbf{#1}}}}

\newcommand{\updownarrows}{{\uparrow\!\downarrow}}

\newcommand{\tri}{\triangledown}
\usepackage{cancel}
\usepackage{nicefrac}
\usepackage{xfrac}
\usepackage{textcomp}

\usepackage{bbold}

\begin{document}
\title{Localizing Transitions via Interaction-Induced Flat Bands}
\author{Alireza Parhizkar and Victor Galitski}
\affiliation{Joint Quantum Institute, Department of Physics, University of Maryland, College Park 20742}
%\date{\today}

\begin{abstract}
This paper presents a theory of interaction-induced band-flattening in strongly correlated electron systems. We begin by illustrating an inherent connection between flat bands and index theorems, and presenting a generic prescription for constructing flat bands by periodically repeating local Hamiltonians with topological zero modes. Specifically, we demonstrate that a Dirac particle in an external, spatially periodic magnetic field can be cast in this form. We derive a condition on the field to produce perfectly flat bands and provide an exact analytical solution for the flat band wave functions. Furthermore, we explore an interacting model of Dirac fermions in a spatially inhomogeneous field. We show that certain Hubbard-Stratonovich configurations exist that ``rectify'' the field configuration, inducing band flattening. We present an explicit model where this localization scenario is energetically favorable -- specifically in Dirac systems with nearly flat bands, where the energy cost of rectifying textures is quadratic in the order parameter, whereas the energy gain from flattening is linear. In conclusion, we discuss alternative symmetry-breaking channels, especially superconductivity, and propose that these interaction-induced band-flattening scenarios represent a generic non-perturbative mechanism for spontaneous symmetry breaking, pertinent to many strongly-correlated electron systems.
\end{abstract}

\date{\today}

\maketitle

\textit{Introduction} -- Flat bands are remarkable phenomena in which electrons cease to propagate  -- their group velocity vanishes: $\partial E(\vect{k})/\partial \vect{k} = 0$. Hence for electrons in a flat band all other energy scales  become relevant as they are now infinitely dominant over the kinetic energy. This makes the flat band a crucial subject of study in strongly correlated electron systems, where the comparison between interaction and kinetic scales often distinguishes distinct phases of matter~\cite{BalentsFlatBand}. In the case of a conventional quadratic dispersion relation, the notions of ``infinite mass'', $m\rightarrow \infty$, and ``flat band'' can be used interchangeably. If the mass divergence occurs as a result of tuning the parameters of an interacting system, then it can be seen as a phase transition. In fact, there has been a long lasting endeavor~\cite{Nozieres1992,VolovikNewClass,FermionCondensate,VolovikExotic,ZhangMassDivergence} of exploringsuch scenarios through interaction-induced renormalization of the effective mass. However, such phenomenological or perturbative approaches often come short in reliably describing singularities in strongly correlated systems. Here, we present an alternative approach, rooted in non-perturbative topological arguments, to recognize flat bands and also provide examples where a transition involving localization or band-flattening occurs due to interactions.

The main idea of our work hinges on a connection between the flat bands and the index theorem~\cite{ASIndex,APSIndex}. We propose a generic prescription for constructing flat bands by a periodic continuation of a local Hamiltonian with a zero mode, which grows into a flat band. We show that one explicit realization of such a construction is an electron in  a periodic classical gauge field. For interacting electrons, interactions can be decoupled in terms of a fluctuating order parameter. There exist classical configurations of the spatially inhomogenous order parameter that give rise to flat bands of the corresponding Bogoliubov quasiparticles. Spontaneous formation of such band-flattening configuration corresponds to a new localization mechanism by many-body effects. Whether such a transition occurs and in what channel is determined by energetics of the model. We show however that for systems where bare electrons' band structure is already close to being flat, this scenario is energetically favorable because the energy gain of flattening the bands is linear in the order parameter while the energy cost of such inhomogeneity is quadratic and hence physics is dominated by the former. 

\textit{Index}---Zero modes of a Hamiltonian and its square are in a one-to-one correspondence. Given a Hamiltonian $H$, if in some basis one can write
\begin{align}
    H^2 =  \left[ 
    \begin{array}{cc}
     D^\dagger D& 0  \\
     0 & DD^\dagger
    \end{array} \right] \, ,
    \label{H2}
\end{align}
then the non-zero eigenvalues of $H^2$ in one block have an equal counterpart in the other block, since if $D^\dagger D\ket{\lambda_+} = \lambda^2 \ket{\lambda_+}$ then $(DD^\dagger)D\ket{\lambda_+} = \lambda^2 D\ket{\lambda_+}$ and thus $\ket{\lambda_-} \equiv D\ket{\lambda_+}/\lambda$ is a properly normalized eigenvector of $DD^\dagger$ with the same positive eigenvalue.
This also means that for each energy level there is another with either the same energy or its negative. So a straightforward conclusion is that if the dimensions of $D^\dagger D$ and $DD^\dagger$, which we respectively call $n_+$ and $n_-$, do not match, the bigger matrix must be enlarged by zero eigenvalues. Therefore, if $n_+ - n_- \neq 0$, a zero energy mode necessarily appears.

A straightforward example of the above construction that possesses zero modes would be certain bipartite chains with odd number of sites~\cite{StoneChain}. However, as we incorporate continuous models, differential operators, and gauge fields, the situation becomes more intricate. In this general case,  $n_+ - n_-$ can be defined as the following index,
\begin{equation}
    n_+ - n_- = \Lim{M \rightarrow \infty} \left[ \Tr f\left(\frac{D^\dagger D}{M^2}\right)- \Tr f\left(\frac{D D^\dagger}{M^2}\right) \right] \, ,
    \label{TrDiff}
\end{equation}
where $f(x)$ is a generic function which is equal to $1$ for $x<1$ but goes to zero rapidly (but smoothly) for $x>1$. Given that the number of non-zero eigenvalues of $D^\dagger D$ and $D D^\dagger$ are the same, they do not contribute to the above expression. The remaining value entirely depends on the difference in the number of their zero eigenvalues.

More generally, for any such Hamiltonian there exists an operator, $\Gamma$, which maps the zero-mode subspace onto itself, $\Gamma \{0\} \rightarrow \{0\}$, and has a zero trace on the compliment subspace. Then the trace of that operator exclusively contains information about zero-modes, and can be used as a proxy for a more general index,
\begin{equation}
    \Tr\, \Gamma = \sum_n \bra{n}\Gamma\ket{n} = \sum_{n\in\{0\}} \! \bra{n}\Gamma\ket{n} \, + \cancel{\sum_{n\notin\{0\}} \! \bra{n}\Gamma\ket{n}} \, . \!  
    \label{TrSep}
\end{equation}
If we demand $\Gamma^2 =1$ then the latter condition can be satisfied if $\Gamma$ distinguishes between corresponding eigenstates, $\Gamma \ket{\lambda_\pm}=\pm\ket{\lambda_\pm}$, or if $\Gamma$ maps corresponding eigenstates into each other, $\Gamma\ket{\lambda_\pm} \propto \ket{\lambda_\mp}$, or a combination of the two. Within the zero mode subspace, $\Gamma$ commutes with the Hamiltonian and thus can be diagonalized with $\pm 1$ eigenvalues for each zero mode. So the expression above can be written as,
\begin{align} \label{TrGamma}
    \Tr\, \Gamma &= N_+ - N_- = \Lim{M\rightarrow \infty} \Tr \Gamma f\left(\frac{H^2}{M^2}\right)  \\
    &=\Lim{M\rightarrow \infty} \left[ \Tr P_+ f\left( \frac{P_+ H^2}{M^2}\right) +  \Tr P_- f\left( \frac{P_- H^2}{M^2}\right) \right] \, , \nonumber
\end{align}
where $N_\pm$ are the number of zero energy modes with $\pm 1$ eigenvalues with respect to $\Gamma$, and $P_\pm \equiv \frac{1\pm\Gamma}{2}$. Note that Eq.~\eqref{TrGamma} resembles Eq.~\eqref{TrDiff}, because the function $f(x)$ is just a regulator dealing with the overcounting of high energy modes~\cite{Fujikawa,Fujikawa2004Book,ParhizkarPathInt,NonAbelianBos,ChiralAnomalyInt}. It is worth noting that if we demand $\Gamma^n =1$ instead, then one way to make the last term in Eq.~\eqref{TrSep} vanish is if $\Gamma$ permutes sets of $n$ non-zero eigenstates with each other. In that case we have,
\begin{equation}
    \Tr \Gamma = \sum_{j=1}^n N_j e^{i\frac{2\pi}{n}j} \, ,
\end{equation}
with $N_j$ being the number of zero-modes with an eigenvalue of $e^{i2\pi j/n}$ with respect to $\Gamma$. This is of relevance e.g. for systems with $C^n$ symmetry such as graphene-based materials. %There are in principle many $\Gamma$s which if come out with non-zero trace lead to zero-modes.

%We can use the method above to find zero modes in the Hamiltonian (or its subspaces) of a generic system. Next we look for flat bands and provide  explicit examples of this construction.

\textit{Flat bands}---To construct an entire flat band we need a set of zero modes and another quantum number to distinguish between all of these zero energy eigenfunctions. If our Hamiltonian has a zero mode confined to a region then we can simply repeat it spatially to obtain a system with discrete translational symmetry. The new system has the lattice momentum as its extra quantum number, and since the zero modes are confined to each ``cell'', we will have an entire band at zero energy.

Consider as a simple example a bipartite lattice where the two sublattices contain different number of sites while the sites of each sublattice are disconnected~\cite{SutherlandLocalization}. The local Hamiltonian describing the bonds of sites inside a cell is given by $H_0$,
\begin{align}
    H_0 = \!  \left[ 
    \begin{array}{cc}
     \mathbf{0}_{N\times N} & D^\dagger  \\
     D & \mathbf{0}_{M\times M}
    \end{array} \right] 
    \! \rightarrow \! \left[ 
    \begin{array}{cc}
     \mathbf{0}_{N\times N} & D^\dagger_\vect{k}  \\
     D_\vect{k} & \mathbf{0}_{M\times M}
    \end{array} \right] \! = H_\vect{k}
     \, , \! 
    \label{Bipartite}
\end{align}
which conveys the fact that in a cell with $M+N$ sites, $N$ sites of one sublattice are connected to $M$ sites of the other sublattice and vise versa, while there is no bond between the sites within each sublattice. Using the argument of the previous section, whenever $N \neq M$, we have $\Tr\Gamma \neq 0$ for $\Gamma=\mathbf{1}_{N\times N}\otimes\mathbf{1}_{2\times 2} - \mathbf{1}_{2\times 2} \otimes \mathbf{1}_{M\times M}$, and therefore there is a zero-mode confined to a cell. We can introduce a lattice momentum simply by letting the components of $D$ depend on Bloch phases given by lattice vectors, promoting $H_0$ to a momentum dependent Hamiltonian $H_\vect{k}$ in Eq.~\eqref{Bipartite}. This readily generates a band structure with a $|N-M|$-fold degenerate flat band at zero energy because the system has a well-defined continuous quantum number (lattice momentum) and independent of the value of that quantum number, there is always $|N-M|$ zero eigenvalues~\footnote{If $\vect{a}_j$ are the lattice vectors, then if the $d_{n,m}$ component of $D^\dagger_{N\times M}$ is proportional to $e^{i\vect{k}\cdot \vect{a}_j}$ the corresponding bond is connecting the $n$-th site of the cell to the $m$-th site of its neighboring cell; the one to the $\vect{a}_j$ neighborhood of the original cell. In the language of periodic boundary condition, these components let the cell return to itself along the $\vect{a}_j$ direction.}. Notice how this argument is independent of the exact content of $D_\vect{k}$ and thus is topological---a smooth change in $D_\vect{k}$ does not alter the flat band.

The previous example dealt with a tight binding model. Let us now introduce differential operators in a continuous model for our next example of a spin-$\frac{1}{2}$ fermion in a periodic magnetic field. Consider a fermion moving across the $x-y$ plane influenced by an inhomogeneous magnetic field applied perpendicularly to the plane along the $\hat z$ direction~\cite{Aharonov}. We can investigate either an electron with a quadratic dispersion, given by the Hamiltonian $h^2 \equiv - \left(\vect{\nabla} - ie\vect{A} \right)^2 - eB_z\sigma^z$, or its square root, $h \equiv i \vect{\sigma} \cdot \left(\nabla - ie\vect{A}\right)$, describing electrons  with a linear dispersion (e.g., in graphene and other two-dimensional Dirac materials). Here, $\vect{\sigma}=(\sigma^x,\sigma^y)$ are the Pauli matrices. These two Hamiltonians share the same zero modes, so it is sufficient to only look at the linear problem.

Note that the Hamiltonian is of the same structure as  Eq.~\eqref{Bipartite} and thus its square has the structure of Eq.~\eqref{H2}. What distinguishes between the two blocks of the Hamiltonian square is $\Gamma = \sigma^z$. This is because the spinor indices are separating the two blocks $D^\dagger D$ and $D D^\dagger$, which are infinite dimensional in real space representation. However, depending on how the gauge field is twisting around, there can be a difference between the dimensionality of these two blocks. Looking at $h^2$, this difference, i.e. $\Tr \Gamma = n_+ - n_-$, is coming from the magnetic field (the only disagreement between the two blocks). We can start with a vanishing magnetic field and gradually raise its strength. When the difference of the dimensions becomes an integer, $n_+ - n_- = 1$, a zero mode appears. But when does this happen?

Looking at Eq.~\eqref{TrDiff}, or Eq.~\eqref{TrGamma}, and recalling that the $D^\dagger D$ and $D D^\dagger$ blocks only differ by $\pm B_z$, we can guess that the difference in the dimensionality of the two blocks -- the index -- must be proportional to the magnetic flux. Let us evaluate this explicitly by working out the right hand side of Eq.~\eqref{TrDiff}. We focus on $D^\dagger D$ first 
\begin{widetext}
\begin{align}
    &\Lim{M\rightarrow\infty} \Tr f\left(\frac{D^\dagger D}{M^2}\right) =\Lim{M\rightarrow\infty} \int_x \int_k e^{i\vect{k}\cdot \vect{x}} f\left(\frac{D^\dagger D}{M^2}\right) e^{-i\vect{k}\cdot \vect{x}} = \Lim{M\rightarrow\infty} M^2 \int_x \int_k f\left[ \left( \vect{k} + \frac{ i\vect{\nabla}  + e\vect{A}}{M} \right)^2 - \frac{eB_z}{M^2}\right] =\\
    &   \Lim{M\rightarrow\infty} M^2 \int_x \int_k \left[ f(\vect{k}^2) + f'(\vect{k}^2) \left( 2\vect{k}\cdot\frac{ i\vect{\nabla}  + e\vect{A}}{M} + \frac{ (i\vect{\nabla}  + e\vect{A})^2}{M^2} - \frac{eB_z}{M^2}  \right) + \frac{1}{2}f''(\vect{k^2}) \frac{\left(2i\vect{k}\cdot \vect{\nabla}  + 2e \vect{k}\cdot \vect{A}\right)^2}{M^2} \right] \, ,  \nonumber
\end{align}
\end{widetext}
where we used $\int_x \int_k \ldots = \int d^2 x \int \frac{d^2 k }{(2\pi)^2} \ldots$ for brevity.
The first equality is the definition of the trace  in the plane wave basis. The second equality is obtained by pulling $e^{-i\vect{k}\cdot \vect{x}}$ through the regulator and rescaling $\vect{k}\rightarrow M\vect{k}$. The second line is the expansion of the regulator about $\vect{k}^2$, up to orders linear in $1/M^2$. All higher orders vanish in the limit $M\rightarrow\infty$. $\Tr f\left(\frac{D D^\dagger}{M^2}\right)$ can be obtained  by  changing the sign of the magnetic field $B_z \to - B_z$ in the expression above for $\Tr f\left(\frac{D^\dagger D}{M^2}\right)$. Subtracting the two from each other yields
\begin{equation}
    n_+ - n_- = \frac{1}{2\pi}\int d^2 x \, e B_z \, ,
    \label{Anomaly}
\end{equation}
so we see that the number of zero-modes is indeed given by the flux. The integration above is over all space, but the same condition applies if we exclude the regions where $B_z$ is zero. We can view the zero-modes as confined within $B_z \neq 0$ regions. Consider for example a triangle-shaped region  with a unit flux of $B_z$, and designate it as $\tri$ (we consider $\tri$ for illustration purposes only, any other shape of a magnetic unit cell is equally valid).  If the rest of the space is tiled by similar magnetic regions, we can think of $\tri$ as having periodic boundary conditions (as in our previous example) and hence periodically-repeating the cell gives us a new quantum number labeling the band of zero-modes. Therefore a periodic magnetic field with an integer flux threading each cell, yields a flat band.

This flat band, as reflected in Eq.~\eqref{Anomaly}, has a connection to the chiral anomaly: The path-integral formulation of the system above is given by $ \int \mathcal{D}\bar\psi\mathcal{\psi}\exp \{\int_x \bar\psi i \slashed{D}\psi\}$, with $[\bar{\psi},\psi]$ being independent fermionic fields and $\slashed{D} = i\partial_t - h$. Zero-modes of $\slashed{D}$ are not zero \textit{energy} modes, but instead they are \textit{on-shell} modes. 
The chiral anomaly counts the number of right-handed modes minus the number of left-handed modes and is given by  Eq.~\eqref{TrGamma} but with $\slashed{D}^2$ substituting $H^2$---the chiral anomaly penalises the off-shell modes rather than non-zero energy modes. However, there is an instance where the two coincide. As the system  develops a flat band at some energy level, say by increasing $B_z$, the ``shell'' is going to lie on that energy level so that off-shell and non-zero energy become synonymous. $\slashed{D}$ and $H$ differ only in the time dimension and in fact they coincide exactly when the band becomes flat and the time dimension becomes irrelevant. A transition amplitude from any state $\ket{\xi_i}$ to any other state $\ket{\xi_f}$ within a flat band is independent of time, $\bra{\xi_f} e^{iHt} \ket{\xi_i} \sim \braket{\xi_f}{\xi_i}$, so time can be removed from the path-integral, e.g. by setting $\partial_t \psi =0$. Further, if the gauge field is periodic in tiles of $\tri$, then we can reduce the path-integral over patches of $\tri$. The anomaly for this ``timeless'' path-integral is given exactly by Eq.~\eqref{Anomaly} with the spatial integral covering only $\tri$. Since the chiral anomaly is the Jacobian of the chiral transformation, after a full chiral rotation, $\psi \rightarrow e^{i2\pi\Gamma}\psi=\psi$ and $\bar\psi\rightarrow e^{i2\pi\Gamma} \bar\psi=\bar\psi$, we must have, 
\begin{equation}
    \int_\tri \mathcal{D}\bar\psi\mathcal{D}\psi e^ {\int_x \bar\psi i \slashed{D}\psi} = \! \int_\tri \mathcal{D}\bar\psi\mathcal{D}\psi e^ {\int_x \bar\psi i \slashed{D}\psi} e^ {i2\pi (n_+ - n_-)} \, . \!
    \label{ChiralRot}
\end{equation}
% Is there an i needed behind the slahed D?
where the subscript $\tri$ means that the path-integral is over one patch of $\tri$, and the last exponential is the Jacobian of the chiral transformation with the exponent being the anomaly. But the above means the path-integral, and therefore the flat band, is realizable only when there is an integer flux through $\tri$.

Our example of a continuous model can be mapped to many systems, among which optical lattices, graphene under periodic strain, and moir{\'e} structures readily come to mind. It also has the desirable feature of being exactly solvable. For a magnetic field, which is periodic with lattice vectors $\vect{a}_{1,2}$, the zero energy flat band wave-functions are,
\begin{equation}
    \psi_\vect{k}^+ = \left( \begin{array}{cc}
     \eta_\vect{k} (z) e^{+\phi}\\
     0   
    \end{array}  \right)  \quad \text{and} \quad
    \psi_\vect{k}^- = \left( \begin{array}{cc}
     0\\
     \bar\eta_\vect{k} (\bar{z}) e^{-\phi}   
    \end{array}  \right) \, ,
    \label{FlatBand}
\end{equation}
with $\vect{\nabla}^2\phi \equiv -B_z$ in Coulomb gauge, $z=x + iy$, and
% $\bar z \equiv z^*$,
\begin{align}
    \eta_\vect{k}(z) = \frac{\sum e^{ i\pi \left[ \frac{a_2}{a_1}\left(n+\frac{\vect{k} \cdot \vect{a}_1}{2\pi}\right)^2 + 2 \left(n+\frac{\vect{k} \cdot \vect{a}_1}{2\pi}\right)\left(\frac{z}{a_1}-\frac{\vect{k} \cdot \vect{a}_2}{2\pi}\right)\right]}}{\sum  e^{i\pi \left[ \frac{a_2}{a_1}n^2 + 2 n\frac{z}{a_1} \right]}} \, 
\end{align}
are elliptic functions. The sum runs over $n=\mathbb{Z}$, $a_{1,2}\equiv a^x_{1,2} + i a^y_{1,2}$ and $\bar\eta$ is defined as $\eta$ but with $a_{1,2}$ replaced by $\bar a_{1,2}$. If we exchange $\eta$ and $\bar\eta$ for identity in above equations, we get only two zero-modes, $\psi^\pm_0$. By introducing $\eta$ we are introducing the additional quantum number. Notice that $\eta_\vect{k} (z)$ is only a function of $z$ and thus, since $\partial_{\bar{z}} \eta_{\vect{k}} (z)=0$, it does not alter the zero-mode equation for $\psi^+_\vect{k}$. Since the function $\eta_\vect{k}(z)$, and consequently the wave-function $\psi^+_\vect{k}$, is quasi-periodic with respect to lattice vectors, $\eta_\vect{k}(z+a_{1,2})=\eta_\vect{k} (z) e^{i\vect{k}\cdot \vect{a}_{1,2}}$, this additional quantum number is the lattice momentum. Each $\psi^\pm_\vect{k}$ satisfies $h\psi^\pm_\vect{k}=0$ and has a lattice momentum of $\vect{k}$. If the magnetic field is composed of equally strong positive and negative patches across the space, then the flat band is two-fold degenerate. On the contrary, if for example there is a constant magnetic field background, then $\phi$ becomes unbounded leading to only one of the wave-functions in Eq.~\eqref{FlatBand} being normalizable.

%A close relative of our continuous example above is the bilayer moir{\'e} graphene at magic angle. The flat band developed at the magic angle is discussed comprehensively in Ref.~\cite{TopolCrit}, where it was shown to originate from the underlying periodic moir{\'e} gauge field, which can be viewed as a special case of the above general discussion.

\textit{Interaction-induced localizing transitions}---We now turn our attention to emergence of flat bands due to interactions. As the most straightforward example naturally following from the previous discussion, consider electrons with current-current interactions:
\begin{align} \label{Int}
    I \! &= \!  \int \! \mathcal{D}\bar\psi\mathcal{D}\psi \exp \! \int \! d^3x \! \left[ \bar\psi i \gamma^\mu \partial_\mu\psi + \frac{\lambda^2}{2}\bar\psi\gamma^\mu\psi\bar\psi\gamma_\mu\psi\right] \!  \! \\
    & \! = \! \int \! \mathcal{D}\bar\psi\mathcal{D}\psi \mathcal{D}a \exp \! \int \! d^3x \! \left[ \bar\psi i \gamma^\mu (\partial_\mu -i\lambda a_\mu ) \psi -\frac{1}{2}a_\mu a^\mu \right] \! , \nonumber
\end{align}
where in the second line we have decoupled the currents via a Hubbard-Stratonovich field, which  depends on space and time. We will focus on classical, spatially inhomogenous configurations only in which case the action reduces to the previous situation of a Dirac particle in a magnetic field.  The on-shell value of the Hubbard-Stratonovich field is $a_\mu = \lambda j_\mu \equiv \lambda \bar\psi\gamma_\mu\psi$. Our periodic magnetic field example readily demonstrates the existence of certain configurations of $a_\mu$ that create flat bands. A non-vanishing expectation value of $\left\langle a_\mu(x) \right\rangle $ implies an ordered symmetry-broken phase of electron currents $j_\mu$. We know from our discussion in the previous section that any configuration of periodic $a_\mu(x)$ that satisfies Eq.~\eqref{Anomaly} for each Hubbard-Stratonovich unit cell (e.g., $\tri$ for the triangular cell), generates a flat band of fermions for this particular realization. If we set $a_0=0$ and demand $\int_\tri d^2x \vect{\nabla}\times \langle \vect{a}(x) \rangle = 1$, it would imply a phase of spontaneously generated periodic loop current textures with a flat band for the corresponding Bogoliubov excitations.

However, there is no apriori reason to choose this or any other particular configuration of the order parameter $\lambda \left\langle \bar\psi\gamma_\mu\psi \right\rangle $ among infinitely many possibilities. Ultimately, the relevant configuration(s) are determined by the energetics of the model. We now argue that there exists a simple general class of model where such a band-flattening transition is natural and indeed energetically favorable. 

Let us reintroduce an external magnetic field, $A_\mu(x)$, into the interacting model. Consider specifically the case where the magnetic field strength  is close but not exactly equal to the magic value where the flat band emerges. Denote the corresponding magic gauge field configuration as $\bar{A}_\mu(x)$. The corresponding path-integral reads:
\begin{align} \label{BInt}
   I \! = \!\! \int \! \mathcal{D}\mu \exp \! \int \! \! d^3 x \! \left[ \bar\psi i \gamma^\mu (\partial_\mu -i e A_\mu -i\lambda a_\mu ) \psi -\frac{1}{2}a_\mu a^\mu \right], 
\end{align}
In contrast to Eq.~\eqref{Int}, what $a_\mu$ needs to provide in order to generate a flat band has reduced to what $A_\mu(x)$ has failed to provide. To ``correct'' the gauge field in order to satisfy the constraint~\ref{Anomaly}, the order parameter needs to acquire a small average value such as
$$
\lambda \left\langle \bar\psi(x)\gamma_\mu\psi(x) \right\rangle  = \bar{A}_\mu(x) - A_\mu(x) \, ,
$$
In this case, the  quadratic term $a_\mu a^\mu$ -- the energy cost of such a texture in Eq.~(\ref{Int}) --is disregardable compared to the energy benefit of flattening the band, which depends only linearly on $a_\mu$. Hence, a transition that flattens the band can happen for $A_\mu(x)$ close enough to the magic configuration, $\bar{A}_\mu(x)$. Note that this argument is independent of the particular gauge field configuration or the physical nature of the gauge field. 

Moir\'e systems~\cite{MoireMaterials,SemiconMoireMaterials,BalentsTwistedMulti,Crepel2023,Popov2023,Guerci2023,MoireGravity} provide a natural physical platform where this scenario can happen and in fact such behavior has been reported in experiment~\cite{Flatexp}. In the case of twisted bilayer graphene~\cite{OriginVishwanath,MacDonald,GrapheneWithTwist,Exp2018correlated,Exp2018unconventional,balents,EmEnSc,TopolCrit} moir\'e patterns give rise to emergent classical gauge fields, which play the role of the background gauge field. At magic twist angles, these fields generate flat bands. 
The recipe for having a spontaneous formation of the flat band (potentially involving spontaneous symmetry breaking) is similar to the previous example: We introduce an ``error'' letting the gauge fields be slightly off their magic value or by breaking the chiral symmetry that enables exact flat bands in the first place, as it happens in actual experiment. Then, if the error is small enough it can become spontaneously corrected by an ordered phase which regenerates the flat band\footnote{If the moir{\'e} gauge field or/and external gauge field configuration is far from the magic value, symmetry breaking into a flat band can still occur, but its recognition requires an analysis of the energetics of the specific model.}. A simple example of this scenario would be to keep the twist angle slightly away from its magic value (see Supplemental Material for a discussion of this qualitative example). Another experimentally-relevant scenario of weakly perturbing the flat-band quantization condition may occur due to deformations of the graphene layers and the moir\'e pattern due to lattice relaxation.

Note that the interaction-induced localization described above generally applies for other types of interactions, where band-flattening would happen due to a spontaneous symmetry breaking in a different channel or channels, with the latter corresponding to a co-existence of different quantum phases. In this context, the superconducting Cooper channel is of particular interest. 

As our last and complimentary discussion, let us consider the possibility of a flat band of Bogoliubov quasiparticles in a superconducting phase. Consider the following action of interacting electrons moving across a two-dimensional Dirac material with the superconducting field, $\Delta$, written in the Nambu space as follows
\begin{align} \label{SC}
     S  \! = \! \int \! d^3x \Bigg( \!
     \bar\Psi  i \! \left[ 
    \begin{array}{cc}
     \partial_t - \vect{\sigma} \cdot \vect{\nabla} & -i \vect{\sigma} \cdot \vect{\Delta}  \\
     -i \vect{\sigma} \cdot \bar{\vect{\Delta}}  & \partial_t + \vect{\sigma} \cdot \vect{\nabla}
    \end{array} \right] \! \Psi - \frac{2}{g}\vect{\bar\Delta}\cdot\vect{\Delta} \! \Bigg)  ,
\end{align}
with $\bar\Psi \equiv \left[ \begin{array}{cc} \bar{\psi}_\uparrow & \psi^T_\downarrow \end{array} \right]$, $\Psi \equiv \left[ \begin{array}{cc} \psi_\uparrow & \bar\psi^T_\downarrow \end{array} \right]^T$  and  $\updownarrows$ designating the spin of the electron while each $\psi$ is a two-component spinor with the pseudo-spin designated by $\pm$ superscripts. Integrating over the auxiliary field $\vect{\Delta}=(\Delta_x,\Delta_y)$ gives the interaction term $g \bar\psi_\uparrow^\pm \bar\psi_\downarrow^\mp \psi_\downarrow^\pm \psi_\uparrow^\mp $ back.

There are many configurations of $\{\Delta,\bar\Delta\}$ giving rise to flat bands and obtainable by methods introduced earlier. When there is a flat band we can remove the time elements from Eq.~\eqref{SC}, as discussed above Eq.~\eqref{ChiralRot}. By switching the second and third spinor components of the fermions $\bar\Psi$ and $\Psi$ in Eq.~\eqref{SC}, the Bogoliubov-deGennes Hamiltonian is 
\begin{align}
    -H =  \left[ 
    \begin{array}{cc}
     \mathbf{0} & D^\dagger  \\
     D & \mathbf{0}
    \end{array} \right] \, , \quad
    D = \left[
    \begin{array}{cc}
     -i\partial_{\bar z} & \Delta_{\bar z}  \\
     {\bar\Delta}_{\bar z} & i\partial_{\bar z}
    \end{array} \right] \, ,
    \label{SCHamiltonian}
\end{align}
with $\Delta_{\bar z} \equiv \Delta_x + i \Delta_y$. As in our previous example of an electron in a periodic magnetic field, if $\psi$ satisfies $D \psi =0$, so does $\eta_\vect{k}(z) \psi$. If a confined zero mode is found we can extend it into a  flat band. An interesting ``magic'' superconducting configuration is given by $\Delta_{\bar z} = \alpha \sum e^{-i \vect{q}_j \cdot \vect{r}} e^{i 2\pi j /3} $ and $\bar\Delta_{\bar z} = -\alpha \sum e^{i \vect{q}_j \cdot \vect{r}} e^{i 2\pi j /3} $ whereby the zero-mode problem above maps to that of the chiral model of the twisted bilayer graphene~\cite{OriginVishwanath}, while the sum is over $j\in\{0,1,2\}$ and $\vect{q}_j \equiv - R(2\pi j/3) \hat{y} $ with $R$ being the vector rotation operator and $\alpha = u/2\nu\sin(\theta/2)$ with $\theta$, $\nu$ and $u$ being the twist angle, Fermi velocity, and the interlayer hopping amplitude respectively.
The first Bogoliubov flat band  emerges at $\alpha\approx 1/\sqrt{3}$.

Note that the method of finding flat bands presented here is blind to what type of fields constitute the Hamiltonian, hence, we can reintroduce the background gauge field simply by minimal coupling $\vect{\nabla}\rightarrow\vect{\nabla}-ie\vect{A}$ whereby the system is provided with a chance to flatten the band through mixed configurations of $\vect{\Delta}$ and $\vect{A}$. Just like in the previous example of the density channel, there is no reason to expect that a superconducting transition into a flat band would appear in a generic system. However, this state becomes competitive in the case of heavy electron bands~\cite{heavy}, where it competes as the rectification mechanism with other channels. 

To summarize, this work presents a general theory of constructing periodic-in-space textures that give rise to flat bands. It formulates a simple condition on the periodic gauge fields tied to the index theorem. More importantly, it shows that this condition can be satisfied not only by external perturbations (fields, rotations, deformations, etc) but also due to the spontaneous formation of periodic electronic textures. A simple proof-of-principle argument is presented that this interaction-induced localization is energetically favorable when the electrons are heavy to begin with. The ideas introduced in this paper are only a seed, and there are many natural extensions of relevance to a variety of strongly correlated systems exhibiting a zoo of proximate symmetry broken phases. These include band-flattening superconducting phases, co-existence of magnetism and superconductivity (where more than one order parameter is needed to flatten the bands), spontaneous formation of moir{\'e} textures in bilayer structures, etc. A careful analysis of experimentally relevant situations where symmetry breaking mechanism via interaction-induced flat bands is realized (including moir{\'e} bilayer graphene and heavy-fermion systems) will be presented elsewhere.

\acknowledgements This work was supported by the National Science Foundation under Grant No. DMR-2037158, the U.S. Army Research Office under Contract No. W911NF1310172, and the Simons Foundation. V. G. also acknowledges support from the National Science Foundation (QLCI grant OMA-2120757). The authors wish to thank Andrey Grankin for useful discussions.

\bibliographystyle{apsrev4-2}
\bibliography{main}

\end{document}

\title{Localizing Transitions via Interaction-Induced Flat Bands
\\
Supplemental Material}
 
\author{Alireza Parhizkar and Victor Galitski}
    \affiliation{Joint Quantum Institute, Department of Physics, University of Maryland, College Park 20742}

\maketitle

\tableofcontents

\section{Periodic Magnetic Field}

A spin-$\frac{1}{2}$ particle has two components. Thus in $(2+1)$ dimensions its hamiltonian is given by $h^2$ if its dispersion is quadratic and $h$ if it is linear, while $h$ is given by the following:
\begin{equation}
    h \equiv i \sigma^j \left(\partial_j - ieA_j\right)
\end{equation}
with $j=\{x,y\}$. For ease of mind we should note that  $h^2$ can also be written in the familiar form below~\cite{Aharonov}:
\begin{align}
    h^2 &\equiv - \left(\vect{\nabla} - ie\vect{A} \right)^2 - eB_z\sigma^z \\
        & = \left[ i\left(\partial_x - ieA_x\right) + \sigma^z \left(\partial_y - ieA_y\right) \right] \times \left[ i\left(\partial_x - ieA_x\right) - \sigma^z \left(\partial_y - ieA_y\right) \right]
\end{align}
where $B_z \equiv \partial_x A_y - \partial_y A_x$. Therefore, solving the zero-mode problem of the quadratic system is the same as the linear one. The zero-mode equation, $h\psi \equiv h(\psi^+,\psi^-)^T = 0$, for one of the components, $\psi^-$, of the wave-function is given as,
\begin{equation}
    \left[\partial_x - i\partial_y - ieA_x -eA_y\right] \psi^-=0
\end{equation}
or
\begin{equation}
    \left[ (\partial_x - e A_y) -i (\partial_y + eA_x) \right] \psi^-=0
\end{equation}
Therefore $\psi^- = e^{-\phi}$ will be a solution to the above equation if,
\begin{equation}
    \partial_x \phi = -eA_y \quad \text{and} \quad \partial_y \phi = eA_x \quad \text{or} \quad \vect{\nabla}^2 \phi = -B
\end{equation}
where $B$ is the external magnetic field normal to the material plane (see Supplemntary material for more details).
\footnote{It is worth seeing that a gauge transformation of $\vect{A} \rightarrow \vect{A} + \vect{\nabla}\Lambda$ leaves the last equation above unchanged but alters the first two. So the above solution works only in a specific gauge but we can reach that gauge through transforming the fermions by a phase.}
The other component of the wave-function, $\psi^+$, must satisfy,
\begin{equation}
    \left[ (\partial_x + e A_y) +i (\partial_y - eA_x) \right] \psi^+=0
\end{equation}
thus $\psi^+ = e^{+\phi}$ will be the corresponding solution for this component. So the general solution to $h\Psi=0$ is as follows,
\begin{equation}
    h\Psi \equiv h \left( \begin{array}{cc}
     \psi^+ \\
     \psi^-   
    \end{array}  \right)
    = h \left( \begin{array}{cc}
     f_+ (x+iy) e^{+\phi}\\
     f_- (x-iy) e^{-\phi}   
    \end{array}  \right) 
    = h \left( \begin{array}{cc}
     f_+ (z) e^{+\phi}\\
     f_- (\bar{z}) e^{-\phi}   
    \end{array}  \right) = 0
    \label{ZeroMode}
\end{equation}
where we used the fact that $\partial_z f(\bar z) = \partial_{\bar{z}} f(z)=0$, with $z \equiv x+iy$. The functions $f_\pm$ must leave $\Psi$ normalizable. Note that one can always choose either of $f_\pm$ to be zero. This becomes important when $|\phi|$ is asymptotically unbounded. Then one can just eliminate the troubling component while the remaining wave-function will necessarily be bounded for a well-behaved magnetic field $B$. For an example of unbounded $\phi$ one can point to $\phi= \alpha(x-x_0)^2 + \beta(y-y_0)^2$ which gives constant magnetic field of $B=2(\alpha+\beta)$. The fact that in the presence of constant magnetic field one of the components is unbounded means that such magnetic field breaks the degeneracy of the ground state.

So far we have found zero energy eigenstates for two dimensional electrons under inhomogeneous magnetic field. Next we wish to find a whole band at zero energy. To have a band structure we need periodicity, which gives rise to an additional quantum number---the crystal momentum. We can create this periodicity by making the external magnetic field periodic with lattice vectors $\vect{a}_{1,2}$. So what we want is to find a set of wave-functions, $\psi_\vect{k}$, which satisfy two conditions, (i) they have zero energy: $h\psi_\vect{k}=0$, (ii) they have crystal momentum of $\vect{k}$:
\begin{equation}
    \psi_\vect{k}(\vect{r} + \vect{a}_{1,2}) = \psi_\vect{k}(\vect{r}) e^{i\vect{k}.\vect{a}_{1,2}} \, .
    \label{QuasiPeriodicity}
\end{equation}
Condition (i) is satisfied as long as $\psi$ has the same form as in Eq. \eqref{ZeroMode}. In order to satisfy condition (ii) note that $\phi$ has the same periodicity as the magnetic field insured by $\vect{\nabla}^2\phi=-B$. So the quasi-periodicity of the wave-function, Eq. \eqref{QuasiPeriodicity} must come from the functions $f_\pm$~\cite{OriginVishwanath,TataTheta}. Consider the following,
\begin{align}
    \eta_\vect{k}(z) = \frac{\sum_{n\in \mathbb{Z}} \exp i\pi \left[ \frac{a_2}{a_1}\left(n+\frac{\vect{k} \cdot \vect{a}_1}{2\pi}\right)^2 + 2 \left(n+\frac{\vect{k} \cdot \vect{a}_1}{2\pi}\right)\left(\frac{z}{a_1}-\frac{\vect{k} \cdot \vect{a}_2}{2\pi}\right)\right]}{\sum_{n\in\mathbb{Z}} \exp i\pi \left[ \frac{a_2}{a_1}n^2 + 2 n\frac{z}{a_1} \right]}
\end{align}
with $a_{1,2}\equiv a^x_{1,2} + i a^y_{1,2}$. It is left to the reader to see that $\eta(z+a_{1,2})=\eta(z) e^{i\vect{k}\cdot \vect{a}_{1,2}}$. Thus there is potentially a two-fold degenerate flat-band:
\begin{equation}
    \psi_\vect{k}^+ = \left( \begin{array}{cc}
     \eta_\vect{k} (z) e^{+\phi}\\
     0   
    \end{array}  \right) \, , \quad \text{and} \quad
    \psi_\vect{k}^- = \left( \begin{array}{cc}
     0\\
     \bar\eta_\vect{k} (\bar{z}) e^{-\phi}   
    \end{array}  \right) \, .
    \label{FlatBand}
\end{equation}
where $\bar\eta$ is defined as $\eta$ but with $a_{1,2}$ replaced by $\bar a_{1,2}$. The normalizability criteria here reduces to normalizability over a unit cell. However, the flat-bands are not always realizable. The obstacle comes from the chiral anomaly or in other words the Atiyah-Singer index~\cite{ASIndex,APSIndex}. This obstacle is surpassed only when the flux through a unit cell is an integer multiple of the flux quantum.

\section{Bilayer Graphene}

As shown in a previous work~\cite{TopolCrit}, the physics of spinless free electrons residing in a system of bilayer graphene while coupled to an external electromagnetic field $A_j$, can be described by the path-integral $\int\mathcal{D}[\bar{\psi},\psi] \, e^{iS_0}$, where the free action $S_0$ is given by,
\begin{equation}
    S_0 =   \int   d^3x \, \bar{\psi} i\slashed{D}\psi  , \quad     \slashed{D}\equiv \gamma^\mu \left( \partial_\mu -ieA_\mu + i \mathcal{A}_\mu \gamma_5 + i  \mathcal{S}_\mu i\gamma_3 \right)  , 
    \label{FAction3D}
\end{equation}
with $\gamma^\mu$ being the four-by-four gamma matrices. In the path-integral above the four component spinor fields $\bar{\psi}$ and $\psi$ are treated as independent, also $\mu\in\{0,1,2\}$ and $\mathcal{A}_\mu$ and $\mathcal{S}_\mu$ are two emergent gauge field with the following components,
\begin{align} \label{GFields}
    \mathcal{A}_0 &= -\frac{v}{\nu_F }\text{Re}[V(\vec{r})] \, , \quad \mathcal{A}_1  = \frac{u }{2\nu_F }\text{Re} [U(\vec{r})+U(-\vec{r})] \, , \quad   \mathcal{A}_2  =  \frac{u }{2\nu_F }\text{Im} [U(\vec{r})+U(-\vec{r})] \, , \nonumber \\
    \mathcal{S}_0 &= +\frac{v }{\nu_F }\text{Im} [V(\vec{r})] \, , \quad \mathcal{S}_1 \, = \frac{u }{2\nu_F }\text{Im}  [U(\vec{r}) - U(-\vec{r})] \, , \quad \mathcal{S}_2 \, =  \frac{u }{2\nu_F }\text{Re}  [U(-\vec{r})-U(\vec{r})] \, ,
\end{align}
with $\nu_F$, $v$ and $u$ respectively being the Fermi velocity of graphene, $AA$ and $AB$ interlayer hopping amplitudes,
\begin{equation}
    V(\vec r)=\sum_{j=0}^2 e^{-i\vec{q}_j \cdot \vec{r}} \,  , \quad U(\vec r)=\sum_{j=0}^2 e^{-i\vec{q}_j \cdot \vec{r}} e^{i  2\pi j /3 } \, ,
\end{equation}
and $\vec{q}_j$ being the three vectors generating the hexagonal structure of the moir\'e reciprocal lattice of the bilayer graphene. In other words $\vect{q}_j$ are the momentum distance between the corresponding Dirac cones of the two layers. For certain values of $\vec{q}_j$---the magic values---the system forms a flat-bad, at which the other bands are maximally pushed away from the flat-band~\cite{OriginVishwanath}. When the same site, $AA$, hopping is turned off and the only mutual deformations of the layers of graphene are that of uniform twist, for the first magic angle (of twist) we have $\vec{q}_0 \approx -\sqrt{3}u/\nu_F \hat{y}$ while $\vec{q}_1$ and $\vec{q}_2$ are given by successive $2\pi/3$ counter-clockwise rotations. The interlayer hopping energy, $u$, is responsible for creating a gap between the otherwise disengaged Dirac cones of each layer. A larger $u$ is more successful in generating a large gap pushing the band towards flatness. When the first flat-band occurs the energy scale of the moir\'e Brillouin zone, the gap required for flatness, is comparable to that of the interlayer hopping $u \approx \nu_F|\vec{q}_j|/\sqrt{3}$. For the next flat-bands, $\nu_F |\vec{q}_j|$ becomes smaller and smaller as the moir\'e Brillouin zone shrinks. While the interlayer hopping energy must be sufficient for an extended range of $\vec{q}_j$, the obstacle for forming a flat-band comes from the chiral anomaly (see Ref.~\cite{TopolCrit}) and it is removed when the gauge fields satisfy a topological condition, namely being able to fit into a moir\'e cell by having a whole winding number within the cell.

Therefore, if the bilayer graphene is twisted away from the magic angle or if $v$ becomes non-zero, the ``flat''-band ceases to be flat and acquires some curvature. However this could be compensated for by applying external fields~\cite{TopolCrit} and thus ``correcting'' the band into flatness.
Here we ask a similar question about interactions: Can introducing interactions somehow make up for stepping away from flatness?

A distortion of the graphene monolayer behaves as an emergent gauge field~\cite{ElectronicProperties,balents,PhononDrivenSupCon,EmEnSc}. A mutual distortion also disrupts the already existing moir\'e pattern by altering $\vec{q}_j$. Lattice vibrations do both, and hence appearing as a fluctuation to $\mathcal{A}_\mu$, $\mathcal{S}_\mu$ and the background electromagnetic field $A_\mu$. These fluctuations are correlated. The electrical interactions between electrons also appears as a fluctuation to the background magnetic field. Considering various interaction possibilities altogether, let us assume, as a trial model, that the following effective interactions are created after taking everything into account,
\begin{align}
    &S_I =   \int   d^3x \left[ \bar{\psi} i\slashed{D}\psi + \frac{\lambda^2}{2} j^\mu j_\mu + \frac{\lambda_5^2}{2} j^\mu_5 j_\mu^5 + \frac{\lambda_3^2}{2} j^\mu_3 j_\mu^3 \right]   , \nonumber \\
    &\text{with} \quad j^\mu \equiv \bar\psi\gamma^\mu\psi \, , \quad j^\mu_5 \equiv \bar\psi\gamma^\mu\gamma_5\psi \, , \quad \text{and} \quad j^\mu_3 \equiv \bar\psi \gamma^\mu i \gamma_3\psi \, .
    \label{FAction3DI}
\end{align}
The current-current interaction term $j^\mu j_\mu$ represents also the Coulomb interaction, in fact it is the Lorenz invariant version of density-density interaction $j^0 j_0 \equiv \psi^\dagger\psi\psi^\dagger\psi$. The $\lambda$s are the interaction strengths. Since a small distortion can cause a huge alteration to the moir\'e pattern, $\lambda_{3,5}$ can be bigger than $\lambda$. We can decouple the interactions above by Hubbard-Stratonovich fields to get,
\begin{align}
    &S_I =   \int   d^3x \left[ \bar{\psi} i\slashed{D}_g\psi -\frac{1}{2} h_\mu h^\mu -\frac{1}{2} a_\mu a^\mu -\frac{1}{2} s_\mu s^\mu \right] \nonumber \\
    & \text{with} \quad \slashed{D}_g \equiv  \gamma^\mu \big[ \partial_\mu - i(e A_\mu +  \lambda h_\mu) + i (\mathcal{A}_\mu - \lambda_5 a_\mu) \gamma_5 + i  (\mathcal{S}_\mu - \lambda_3 s_\mu) i\gamma_3 \big]  \, , 
    \label{FAction3DDecoup}
\end{align}
where now the path-integral, $\int\mathcal{D}[\bar{\psi},\psi,h_\mu, a_\mu,s_\mu] \, e^{iS_I}$, also goes over the auxiliary fields $h_\mu$, $a_\mu$ and $s_\mu$, and the generalized Dirac operator $\slashed{D}_g$ includes the Hubbard-Stratonovich fields and treats them in the same fashion as their corresponding gauge field. If we wish to keep, for example, only the density-density part of the $j^\mu j_\mu$ interaction, then we only need to keep $a_0$ and remove the rest of the components of the auxiliary fields. Also, the on-shell values are given by $h^\mu \doteq \lambda j_\mu$, $a^\mu \doteq \lambda_5 j_5^\mu$ and $s^\mu \doteq \lambda_3 j_3^\mu$ which upon substitution in above yield Eq. \eqref{FAction3DI} back. (The dotted equality sign conveys that the equality holds on shell.)

The low energy physics of the bilayer graphene is completely governed by the emergent gauge fields $\mathcal{A}_\mu$ and $\mathcal{S}_\mu$, in particular, flat-bands happens for specific configurations of these fields. In the case that we are, for any reason, away from the flat-band, it is clear from Eq. \eqref{FAction3DDecoup} that there can always be found order parameter configurations which takes us back to the perfect flat-band: Let us consider a generic configuration of the fields for which we have a flat-band,
\begin{equation}
    \left\{ A_\mu, \mathcal{A}_\mu , \mathcal{S}_\mu \right\}\big|_{\text{flat-band}}  = \left\{ \bar{A}_\mu , \bar{\mathcal{A}}_\mu , \bar{\mathcal{S}}_\mu \right\} \, ,
\end{equation}
and then imagine the system being slightly away from the flat-band by some errors, $\epsilon_\mu$, occurring in the gauge fields,
\begin{equation}
    \left\{ A_\mu, \mathcal{A}_\mu , \mathcal{S}_\mu \right\}  = \left\{ \bar{A}_\mu + \epsilon^A_\mu, \bar{\mathcal{A}}_\mu + \epsilon^\mathcal{A}_\mu, \bar{\mathcal{S}}_\mu + \epsilon^\mathcal{S}_\mu \right\} \, .
\end{equation}
Then there is always a configuration of the auxiliary fields (among infinitely many) to correct the error and take the band structure back to the flatness,
\begin{equation}
    \left\{ h_\mu, a_\mu , s_\mu \right\}\big|_{\text{flat-band}}  = \left\{-\frac{e}{\lambda}\epsilon^A_\mu, \frac{1}{\lambda_5}\epsilon^\mathcal{A}_\mu, \frac{1}{\lambda_3}\epsilon^\mathcal{S}_\mu \right\} \, .
\end{equation}
According to the criterion discussed in the paper, the above is just one of the many configurations of $\{ h_\mu, a_\mu , s_\mu \}$ that can regenerate flatness, but it is also one of the configurations which are as small as the errors.

An instance of above occurs for the chiral model of the twisted bilayer graphene~\cite{OriginVishwanath} when the twist angle is slightly away from the magic angle. To compensate for this error we only need non-zero $a_{1,2}$ and $s_{1,2}$, and as discussed above there exists a configuration of $a_{1,2}$ and $s_{1,2}$ that both does the job and is small. The electrons are coupled to $\mathcal{A}_\mu -\lambda_5 a_\mu$ and $\mathcal{S}_\mu - \lambda_3 s_\mu$, thus in fact whatever can be known about the whole band structure (including the density of states) is a function of these two terms. Thus for the energy of the whole system at zero temperature, $E$, constituted by the energy of fermions, $\mathcal{E}_{\nu}$, which also depends the chemical potential $\nu$, and the energy cost of the (auxiliary) gauge fields, we have,
\begin{align} \label{EnergyParts}
    E &= \mathcal{E}_{\nu}[\mathcal{A}_\mu -\lambda_5 a_\mu , \mathcal{S}_\mu - \lambda_3 s_\mu] + \frac{1}{2}a_\mu a^\mu + \frac{1}{2} s_\mu s^\mu \\
    &\approx \mathcal{E}_{\nu}\Bigg|_{\mathcal{A}_\mu,\mathcal{S}_\mu} - \lambda_5 \frac{\delta \mathcal{E}}{\delta \mathcal{A}^\mu}\Bigg|_{\mathcal{A}_\mu,\mathcal{S}_\mu} a^\mu - \lambda_3 \frac{\delta \mathcal{E}}{\delta \mathcal{S}^\mu}\Bigg|_{\mathcal{A}_\mu,\mathcal{S}_\mu} s^\mu + \frac{1}{2}a_\mu a^\mu + \frac{1}{2} s_\mu s^
    \mu \, , \nonumber
\end{align}
so for an small enough deviation from flatness (or in other words curvature) and consequently small enough $a_\mu$ and $s_\mu$ a transition to flat-band can happen. For the case of the example at hand, we know that at the magic angle other bands are at a maximum distance from the flat-band~\cite{OriginVishwanath}. As the hopping potential pushes the band to its extreme (prefect flatness) it also maximally pushes the other bands away from it. This means that the linear part of Eq. \eqref{EnergyParts} can have a winning energy benefit for certain values of the chemical potential.

\bibliographystyle{apsrev4-2}
\bibliography{main}